\documentclass[twocolumn,aps,prl,groupedaddress,showpacs]{revtex4}

\usepackage{graphicx}
\usepackage{amsmath}

\begin{document}

\title{Density engineering of an oscillating soliton/vortex ring in a Bose-Einstein condensate}

\author{Itay Shomroni}
\author{Elias Lahoud}
\author{Shahar Levy}
\author{Jeff Steinhauer}
\email[E-mail address: ]{jeffs@physics.technion.ac.il} \affiliation{Technion--Israel Institute of Technology}

\begin{abstract}
When two Bose-Einstein condensates (BEC's) collide with high collisional energy, the celebrated matter wave
interference pattern results. For lower collisional energies the repulsive interaction energy becomes
significant, and the interference pattern evolves into an array of grey solitons. The lowest collisional energy,
producing a single pair of solitons, has not been probed. We use density engineering on the healing length scale
to produce such a pair of solitons. These solitons then evolve periodically between vortex rings and solitons,
which we image in-situ on the healing length scale. The stable, periodic evolution is in sharp contrast to the
behavior of previous experiments, in which the solitons decay irreversibly into vortex rings via the snake
instability. The evolution can be understood in terms of conservation of mass and energy in a narrow condensate.
The periodic oscillation between two qualitatively different forms seems to be a rare phenomenon in nature.
\end{abstract}


\maketitle

\begin{figure}
\includegraphics{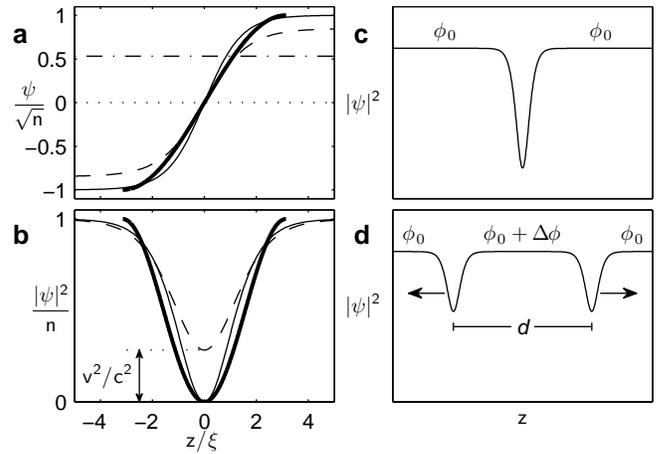}
\caption{Solitons and matter-wave interference fringes. (a)~The order parameter near the transition from
interference fringes to solitons. The bold solid curve depicts an interference fringe with $\lambda=4\pi\xi$.
The narrow solid curve shows a soliton with $v=0$. Both of these order parameters are real. The dashed
(dash-dotted) curve indicates the real (imaginary) part of a grey soliton, with $v=0.53c$. (b)~The density
associated with the curves of part~(a). (c)~The initial density and phase of our experiment, created by density
engineering. (d)~The density and phase of two counterpropagating solitons.\label{fig1}}
\end{figure}

We consider two condensates separated in space, and given an initial collisional energy $E_\text{col}$, which
could be in the form of kinetic energy, or potential energy in a magnetic trap.  The condensates are then
allowed to collide in the $z$ direction. For negligible interactions, the two colliding condensates will
interfere like two non-interacting plane waves, forming a standing wave interference pattern with wavefunction
$\psi\propto\cos(kz)$, where the wavelength of the pattern is $\lambda=2\pi/k$, and $k$ is given by
$\hbar^2k^2/2m=E_\text{col}/N$, where $m$ is the atomic mass and $N$ is the number of atoms. The density of the
pattern is therefore of the form~\cite{ref1,ref2} $n=|\psi|^2\propto\cos^2(kz)$. The initial energy
$E_\text{col}$ supplies the kinetic energy of the standing wave, proportional to $\partial^2\psi/\partial z^2$,
as well as the negligible interaction energy $N\mu/4$. However, if $E_\text{col}$ is decreased to become
comparable to $N\mu/4$, the form of the fringes will be modified by interactions. It will become energetically
favorable to form an array of grey solitons, rather than a cosine-squared pattern~\cite{ref2}. The criterion
$E_\text{col}<N\mu/4$ can be written as $\lambda > 4\pi\xi$, where $\xi$ is the healing length, the approximate
width of a grey soliton~\cite{ref13,ref14}. In other words, as $E_\text{col}$ decreases, the wavelength of the
standing wave interference pattern increases until the wavelength reaches the size of a soliton. At this point,
the fringes will evolve into solitons. A cosine-squared fringe and a soliton are actually very similar in form,
as seen in Fig.~\ref{fig1}a,~b. As $E_\text{col}$ decreases further, the number of solitons will decrease,
maintaining conservation of energy~\cite{ref3}. The lowest collisional energies will result in two grey solitons
traveling in opposite directions. Such a low collisional energy is achieved in our experiment by initially
separating the two condensates by a barrier whose width is on the order of $\xi$, as shown in Fig.~\ref{fig1}c.
The energy of this initial configuration relative to the ground state is on the order of
$E_\text{col}=N\mu(\xi/L)$, where $L$ is the length of the condensate in the $z$ direction. This is comparable
to the energy of a grey soliton computed in one dimension, which is on the order of
\begin{equation}
E_\text{sol} = N\mu(\xi/L)\left(1-v^2/c^2\right)^{3/2}, \label{eq1}
\end{equation}
where $v$ is the subsonic soliton velocity, and $c$ is the speed of sound in the condensate~\cite{ref13}.
$E_\text{col}$ is thus sufficient to create a pair of solitons.

Solitons can also be created by phase engineering~\cite{ref15,ref16}. There is a phase step $\Delta\phi\leq\pi$
across the grey soliton, given by $\Delta\phi = 2\cos^{-1}(v/c)$. By imposing this phase step, the density
profile will evolve into that of a soliton. We can understand this process by considering the current density
$\mathbf{J} = (\hbar/m)n\nabla\phi$. By imposing the phase step, this current density is significant in the
region of the step only. This corresponds to a divergence in the flow field, $\nabla\cdot\mathbf{J} = \partial
J_z/\partial z \neq 0$. Since the condensate obeys the continuity equation $\nabla\cdot\mathbf{J} = -\partial
n/\partial t$, The divergence depletes the density, forming the soliton minimum.

We use density engineering to form solitons, a complementary technique to phase engineering. We impose the
density profile, and the phase evolves into that of a soliton. The process is illustrated in Fig.~\ref{fig1}d
which shows two solitons separated by a distance $d$. They are moving in opposite directions, with the
corresponding opposite phase steps. Our initial configuration shown in Fig.~\ref{fig1}c can be thought of as the
$d=0$ case of Fig.~\ref{fig1}d. Our initial configuration thus corresponds to two overlapping,
counterpropagating solitons.

\begin{figure}
\includegraphics{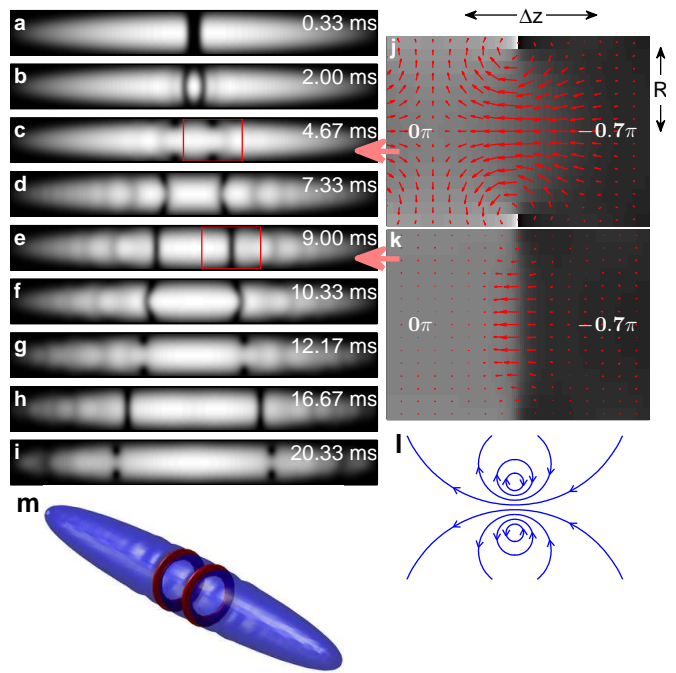}
\caption{Simulation of the periodic soliton/vortex ring. This simulation shows $3\times 10^4$ atoms.
(a)--(i)~Slices of the condensate. The shades of grey indicate the density.  (a)~The condensate during the
turn-off of the barrier. (b), (e), (h)~The soliton stage of the periodic soliton/vortex ring.  (c), (g), (i)~The
vortex ring stage. The axis of the 2 counterpropagating vortex rings is horizontal in the image. The slice of
the vortex ring appears as two dark spots, separated vertically. (d), (f)~The transitions between the soliton
and vortex ring stages. (j)~The shades of grey indicate the phase, and the red vectors indicate $\mathbf{J}$.
The image is an enlargement of the area between the red lines of part~(c). The values $0$ and $-0.7\pi$ indicate
the phase on either side of the phase step.  (k)~As in part~(j), except that the enlargement refers to part~(e).
(l) An illustration of the slice of a vortex ring in an infinite, homogeneous condensate.  (m)~An illustration
of the 2 counterpropagating vortex rings of part~(c), showing a surface of constant density (blue) as well as a
surface of constant flow speed (red).\label{fig2}}
\end{figure}

In a one-dimensional (1D) condensate, solitons are stable objects. In three dimensions however, grey solitons
were seen to decay irreversibly into vortex rings via the snake
instability~\cite{ref4,ref5,ref6,ref7,ref8,ref9,ref10,ref11}, which was originally measured in optical
solitons~\cite{ref12}. We find a very different, stable behavior shown in Fig.~\ref{fig2} by a simulation of the
3D Gross-Pitaevskii equation (GPE)~\cite{ref13}. The soliton evolves into a vortex ring, which subsequently
evolves back into a soliton. This process repeats itself many times with a period of 7--8~ms. We call this a
``periodic soliton/vortex ring''. The periodic soliton/vortex ring is stable, surviving the reflection from the
end of the condensate, as well as the subsequent collision in the middle of the condensate. The periodic
soliton/vortex ring has a constant phase step ($0.7\pi$ for our experiment), as shown in Fig.~\ref{fig2}j,~k.

First, we will understand the evolution of the soliton into a vortex ring in terms of conservation of energy.
Due to the higher density near the center of the condensate, the soliton moves fastest in this
region~\cite{ref16}. This causes the soliton to curve~\cite{ref16}, as shown in Fig.~\ref{fig2}f. Due to the
curvature, the area of the soliton has increased, corresponding to an increase in the soliton energy. To
conserve the energy of the soliton, the increase in area must be compensated by an increase in $v$, by
equation~\ref{eq1}. This corresponds to a reduction in the depth of the soliton, as shown in Fig.~\ref{fig1}b.
The soliton thus moves with ever-increasing speed and decreasing depth, until the density minimum vanishes
altogether, leaving a vortex ring, as shown in Fig.~\ref{fig2}c,~g, and~i. The appearance of the vortex ring
does not violate Kelvin's theorem (conservation of circulation), because the vortex core can enter and exit the
condensate via the nearby boundary~\cite{ref17}.

The fate of vortex rings in a quantum fluid is a long-standing field of study~\cite{ref17,ref18}. In superfluid
$^4$He at non-zero temperatures, it was suggested that the radius of a vortex ring shrinks in time due to
dissipation, until it evolves into a roton. We can understand the very different evolution of our vortex ring
into a soliton in terms of conservation of mass. We consider a vortex ring in an infinite homogeneous
condensate~\cite{ref13,ref19}, as illustrated~\cite{ref18} in Fig.~\ref{fig2}l. The density is constant in time,
maintained by the zero divergence of the flow ($\nabla\cdot\mathbf J = 0$). In other words, after the flow
traverses the center of the vortex ring from right to left, it returns to the right outside the vortex ring. In
contrast, our vortex ring is close to the edge of the condensate, as shown in Fig.~\ref{fig2}c,~g, and~i. The
flow lines therefore have no return path to the right side, causing $\nabla\cdot\mathbf J \neq 0$. The
divergence depletes the density, forming the soliton minimum. This situation is very similar to that of phase
engineering of a condensate described above. In fact, the vortex ring has a phase step, as shown in
Fig.~\ref{fig2}j. The phase step occurs on a length scale $\Delta z$ on the order of the radius of the vortex
ring, which is also the radius $R$ of the condensate. Thus, for creating solitons from vortex rings, $R$ should
not be too much larger than $\xi$. This can also be understood energetically. The periodic soliton/vortex ring
requires that the soliton and the vortex ring have the same energy. The ratio of these energies contains a
factor of $R/\xi$, placing an upper limit on $R$. By the GPE simulation, we find that the vortex ring evolves
into a soliton for $R\leq 13\xi$. This requirement implies that the number of atoms in the condensate should not
be too large. For our trap frequencies for example, simulations show the periodic soliton/vortex ring for $N\leq
7\times 10^4$.

\begin{figure}
\includegraphics{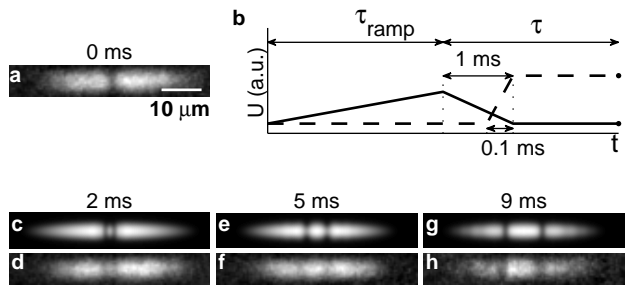}
\caption{In-situ images of the vortex rings and solitons. The experimental parts of the figure show averages of
2 to 6 destructive phase contrast images, employing small detuning. The condensates contain $1\times 10^4$
atoms.  (a)~The initial condition with the barrier on.  (b)~The height of the barrier as a function of time (not
to scale). The solid curve corresponds to density engineering of solitons. The dashed curve corresponds to the
creation of sound pulses.  (c)~Simulation of the initially formed pair of solitons at $\tau=2\,\mathrm{ms}$.
This is an integrated image which includes finite resolution and depth-of-field.  (d)~In-situ image of the two
counterpropagating solitons at 2~ms.  (e)~Simulation of the vortex ring stage at $\tau=5\,\mathrm{ms}$. This is
an integrated image which includes finite resolution and depth-of-field.  (f)~In-situ image of two
counterpropagating vortex rings at 5~ms.  (g)~Simulation of the soliton stage at $\tau=9\,\mathrm{ms}$. This is
an integrated image which includes finite resolution and depth-of-field.  (h)~In-situ image of two
counterpropagating solitons at $\tau=9\,\mathrm{ms}$.\label{fig3}}
\end{figure}

\begin{figure}
\includegraphics{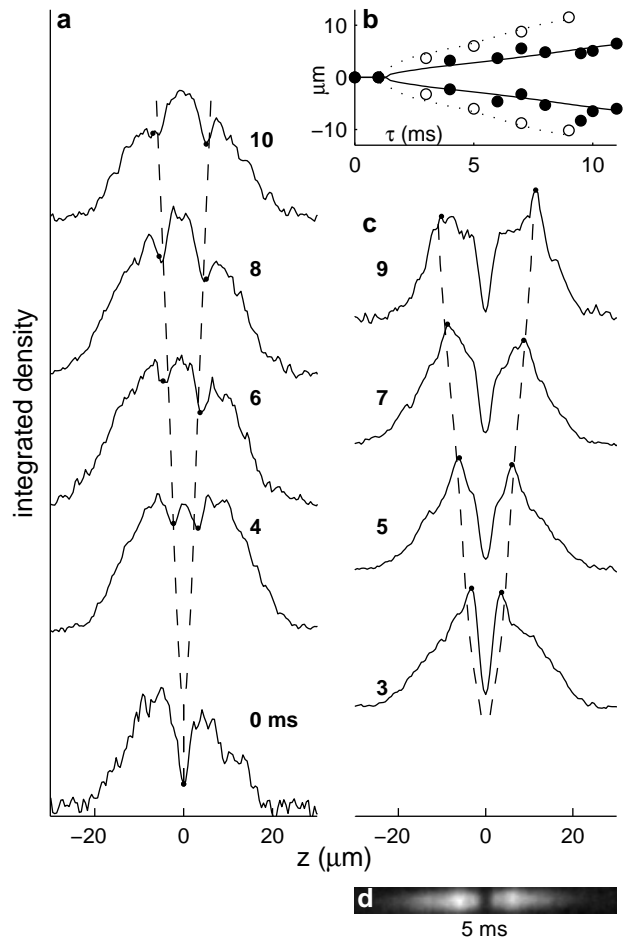}
\caption{The speed of the periodic soliton/vortex ring. The condensates contain $3\times 10^4$ atoms.
(a)~Integrated profiles of the periodic soliton/vortex ring. Each profile is an average of several images,
except for the 0~ms profile.  The dashed curves are the result of the simulation.  (b)~Closed circles indicate
the position of the periodic soliton/vortex ring as a function of time. Open circles indicate the position of
the sound pulses. The solid and dotted curves are simulations of the periodic soliton/vortex ring and sound
pulses, respectively.  (c)~The sound pulses. Each profile is an average of several images. The dashed curves
indicate the simulation.  (d)~Two counterpropagating sound pulses at 5~ms.\label{fig4}}
\end{figure}

Our experimental apparatus is basically described in Ref.~\cite{ref20}. After reaching BEC, the trap frequencies
are adiabatically reduced by half in 350~ms to 117~Hz and 13~Hz in the radial and axial directions,
respectively. The resulting condensate contains $1\times 10^4$ or $3\times 10^4$ $^{87}$Rb atoms, as specified
in the captions of Figs.~\ref{fig3} and~\ref{fig4}. The condensate with $3\times 10^4$ atoms has $\mu/h =
580\,\text{Hz}$ ($h$ is Planck's constant), corresponding to a minimum healing length of $\xi=0.32\,\mathrm{\mu
m}$, which is found at the center of the condensate. The healing length is larger at the lower densities found
away from the center. The radii of this condensate are $89\xi$ and $R=10\xi$ in the axial and radial directions
respectively. The condensate with $1\times 10^4$ atoms has even larger $\xi$ and smaller $R$.

Density engineering is performed by a laser beam, blue detuned by 6~nm from the 780~nm resonance. The condensate
is split in the radial direction by the highly elongated laser beam with an axial diameter of $4.4\xi$
($1.4\,\mathrm{\mu m}$). By the 1D GPE simulations of Ref.~\cite{ref3}, as well as our 3D GPE simulations, this
diameter of roughly $4\xi$ is the maximum which will create a single pair of solitons.

We adiabatically ramp up the laser barrier in $\tau_\mathrm{ramp} = 200\,\mathrm{ms}$ to a height $U =
h(3000\,\mathrm{Hz})$, creating a BEC with a density minimum, as shown in Fig.~\ref{fig3}a. The potential is
then rapidly turned off in 1~ms, resulting in the initial condition illustrated in Fig.~\ref{fig1}c. This
sequence is indicated by the solid curve in Fig.~\ref{fig3}b. As seen in Fig.~\ref{fig3}d, one can distinguish
the two counterpropagating solitons after 2~ms, which result from the interference between the two condensates
of Fig.~\ref{fig3}a, colliding with low energy. Figure~\ref{fig3}d corresponds to Fig.~\ref{fig1}d.
Figure~\ref{fig3}c shows the simulation at 2~ms with imaging effects, including integration of the density
perpendicular to the image, finite resolution, and finite depth of field. The experiment (Fig.~\ref{fig3}d) is
seen to agree well with the simulation (Fig.~\ref{fig3}c).

After a time $\tau=5\,\mathrm{ms}$, we obtain in-situ images of the pair of counterpropagating vortex rings of
the periodic soliton/vortex ring, as shown in Fig.~\ref{fig3}f. In the image, the two vortex rings are seen as
weak density minima. Again, good agreement is seen with the simulation of Fig.~\ref{fig3}e. After each of the
two vortex rings has evolved back into a soliton, we image the soliton stage at $\tau=9\,\mathrm{ms}$ as shown
in Fig.~\ref{fig3}h. Two strong, slit-like solitons are seen. We find that the solitons are more visible for the
small atom number shown in the figure. Good agreement with the simulation of Fig.~\ref{fig3}g is seen.

In order to measure the speed of the periodic soliton/vortex ring, we observe its position as a function of
time, as shown in the integrated profiles of Fig.~\ref{fig4}a, and the filled circles of Fig.~\ref{fig4}b. We
see that the speed (the slope) is approximately constant in time, and is therefore the same for the soliton
stage and the vortex ring stage. The speed is observed to be $v = 0.66\pm 0.08\,\mathrm{mm\,s^{-1}}$, for
$N=3\times 10^4$. The simulation gives the same results qualitatively and quantitatively, as indicated by the
dashed and solid curves of Fig.~\ref{fig4}a,~b. This speed is of the same order of magnitude as the speed
$v\approx (\hbar/2mR)\ln(1.59R/\xi)\approx 0.3\,\mathrm{mm\,s^{-1}}$ predicted for a vortex ring in an infinite,
homogeneous condensate~\cite{ref13,ref17,ref18}.

The speed of a grey soliton is less than $c$, so it is interesting to compare the speed of the periodic
soliton/vortex ring to $c$. We measure $c$ by the technique of Ref.~\cite{ref21}. Specifically, we rapidly turn
on the barrier in $100\,\mathrm{\mu s}$ and leave it on as indicated by the dashed curve of Fig.~\ref{fig3}b,
creating two counterpropagating sound pulses, as shown after 5~ms in Fig.~\ref{fig4}d. The profiles of
Fig.~\ref{fig4}c and the open circles of Fig.~\ref{fig4}b show the position of the sound pulses as a function of
time, giving $c=1.24\pm 0.07\,\mathrm{mm\,s^{-1}}$, which agrees with the theoretical value~\cite{ref22} of
$c=\sqrt{\mu/2m} = 1.15\pm 0.09\,\mathrm{mm\,s^{-1}}$. The dotted and dashed curves of Fig.~\ref{fig4}b,~c show
the results of the GPE simulation, which agree well with the experiment. We thus find that the periodic
soliton/vortex ring moves slower than the speed of sound, at $v=0.53c$.

In conclusion, we have achieved density engineering of a pair of counterpropagating grey solitons, which are the
result of a low-energy collision between 2 BEC's. The close relationship between solitons and matter wave
interference fringes elucidates the quantum mechanical nature of solitons in a BEC. Due to the inhomogeneous
nature of the narrow condensate, each soliton evolves into a periodic soliton/vortex ring. We explain this
evolution in terms of conservation of mass and energy. Since vortex rings and solitons can be considered to be
quasiparticles~\cite{ref7,ref13,ref17}, perhaps the periodic soliton/vortex ring could additionally be described
as a Rabi oscillation between these quasiparticle states. Such a Rabi oscillation is seen in optics, between two
soliton states~\cite{ref23}. As a macroscopic non-linear entity, the periodic soliton/vortex ring has the
unusual property that it oscillates between two qualitatively different forms. This presents a puzzle to all
branches of science, to categorize this effect by finding other systems with an oscillating nature.


\begin{acknowledgments}
We thank Avy Soffer, Moti Segev, Wolfgang Ketterle, Anna Minguzzi, and Roee Ozeri for helpful discussions. This
work was supported by the Israel Science Foundation.
\end{acknowledgments}

\bibliography{soliton_vr}

\end{document}